\documentclass[preprint,showpacs,preprintnumbers,amsmath,amssymb]{revtex4}

\usepackage{graphicx}
\usepackage{dcolumn}
\usepackage{bm}

\begin{document}

\title{Spontaneous Symmetry Breaking in General Relativity.\\Brane World Concept.}

\author{Boris E. Meierovich}
 
\email{meierovich@mail.ru}
\affiliation{P.L.Kapitza Institute for Physical Problems  \\
2 Kosygina str., Moscow 119334, Russia}

 \homepage{http://www.kapitza.ras.ru/people/meierovich/}

\date{\today}

\begin{abstract}
Gravitational properties of a hedge-hog type topological defect in
two extra dimensions are considered in General Relativity
employing a vector as the order parameter. The developed macroscopic theory of phase transitions with spontaneous symmetry breaking is applied to the analysis of possible "thick" brane structures.  The previous
considerations were done using the order parameter in the form of
a multiplet in a target space of scalar fields. The difference of
these two approaches is analyzed and demonstrated in detail.

There are two different symmetries  of regular solutions of  Einstein equations for a hedgehog type vector order parameter. Both solutions are analyzed in parallel analytically and numerically. Regular configurations in cases of  vector order parameter have one more free parameter in comparison with the scalar multiplet solutions.

It is shown that the existence of a
negative cosmological constant is sufficient for the spontaneous
symmetry breaking of the initially plain bulk.

Regular
configurations have a growing gravitational potential and are able
to trap the matter to the brane. Among others there are solutions with the gravitational potential having several
points of minimum. Identical in the uniform bulk spin-less particles, being trapped within
separate points of minimum, acquire different masses and appear to an
observer within the brane as different particles with integer spins.
\end{abstract}

\pacs{04.50.+h, 98.80.Cq}
\maketitle

\section{\label{sec:level1}Introduction} 
The theories of brane world and multidimensional gravity are widely
discussed in the literature. They continue numerous attempts to find the origin of enormous hierarchy of energy and mass scales observed in nature, to explain the dark matter and dark energy effects, and other long-standing problems in physics. One can find a great variety of separate brane-world models in the literature: thin and thick branes in five and more dimensions, flat or curved branes and bulk, different mechanisms of brane formation, etc.

From my point of view the most natural approach is to consider the appearance of a
distinguished hyper-surface in the space-time manifold as a result of a phase transition with spontaneous
symmetry breaking in the early phase of the Universe formation. The macroscopic Landau theory of phase transitions with spontaneous symmetry breaking allows to consider the brane world concept self-consistently, even without the knowledge of the nature of physical vacuum. It is reasonable to regard the brane world as a topological defect, which inevitably appears as a result of a phase transition with spontaneous symmetry breaking.

The properties of topological defects (strings,  monopoles, ...) are generally described in general relativity with the aid of a multiplet of scalar fields
forming a hedgehog-type configuration in extra dimensions (see \cite{Bron 1} and references there in). The scalar multiplet plays the role of the order parameter. The hedgehog-type multiplet configuration is proportional to a unit vector in the Euclidean target
space of scalar fields. Though this model is self-consistent, it is not the direct way for generalization of a plane monopole to the curved space-time.

In a flat space-time there is no difference between a vector and a hedgehog-type multiplet of scalar fields. On the contrary, in a curved space-time scalar multiplets and real vectors are transformed differently. For this reason in general relativity the two
approaches (a multiplet of scalar fields and a vector order
parameter) give different results which are worth to be compared. It is the subject of this paper. A priory it seems more difficult to deal with a vector order parameter, and, probably, it is the reason why I couldn't find in the literature any papers considering phase transitions with a hedgehog-type vector order parameter in general relativity.

The spontaneous symmetry breaking with a hedgehog-type vector order parameter in application to the brane world with two extra dimensions is considered in this paper. General formulae (section \ref{General approach}) are applied to the analytical (section \ref{Global string in extra dimensions}) and numerical (\ref{Numerical analysis}) analysis of gravitational properties of global strings in two extra dimensions. The results are summarized in section \ref{Concluding remarks}.

\section{\label{General approach} General approach}

\subsection{\label{Lagrangian}Lagrangian}

The order parameter enters the Lagrangian via scalar bilinear
combinations of its covariant derivatives and via a scalar potential $V$
allowing the spontaneous symmetry breaking. If $\phi _{I}$ is a
vector order parameter, then $V$\ should$\ $be$\ $a function of
the scalar $\phi ^{K}\phi _{K}=g^{IK}\phi _{I}\phi _{K},$ and a
bilinear combination of the derivatives is a tensor
\begin{equation*}
 S_{IKLM}=\phi_{I;K}\phi_{L;M}.
\end{equation*}
Index $_{;K}$ is used as usual for covariant derivatives. There
are three ways to simplify $S_{IKLM}$ into scalars, so the most
general form of the scalar $S,$ formed via contractions of
$S_{IKLM},$ is
\begin{equation}
S=A\left(
\phi_{;K}^{K}\right)^{2}+B\phi_{;K}^{L}\phi_{L}^{;K}+C\phi
_{;K}^{M}\phi _{;M}^{K}, \label{Scalar S}
\end{equation}
where $A,B,$ and $C$ are arbitrary constants. Different
topological defects can be classified by these parameters. In
curved space-time the scalar $S$ depends not only on the
derivatives of the order parameter, but also on the derivatives of
the metric tensor. This is the principle difference between a
vector and a multiplet of scalar fields. \qquad \qquad

The general form of the Lagrangian determining gravitational
properties of topological defects with a vector order parameter is
\begin{equation}
 L\left(\phi_{I},g^{IK},\frac{\partial g_{IK}}{\partial
x^{L}}\right)=L_{g}+L_{d},\notag
\end{equation}
where
\begin{equation}
 L_{g}=\frac{R}{2\kappa^{2}},
 \end{equation} 
 \begin{equation}
L_{d}=A\left( \phi _{;K}^{K}\right)
^{2}+B\phi _{;K}^{I}\phi _{I}^{;K}+C\phi _{;K}^{I}\phi
_{;I}^{K}-V\left( \phi ^{K}\phi _{K}\right) .\label{Ldef}
\end{equation}
$L_{g}$ is the Lagrangian of the gravitational field, $R$ is the
scalar curvature of space-time, $\kappa ^{2}$ is the
(multidimensional) gravitational constant, and $L_{d}$ is the
Lagrangian of a topological defect. Covariant derivation
\begin{equation} \phi _{P;M}=\frac{\partial \phi
_{P}}{\partial x^{M}}-\frac{1}{2}g^{LA}\left( \frac{\partial
g_{AM}}{\partial x^{P}}+\frac{\partial g_{AP}}{\partial
x^{M}}-\frac{\partial g_{MP}}{\partial x^{A}}\right) \phi _{L}
\end{equation}
and razing of indexes $\phi ^{K}=g^{IK}\phi _{I}$ contain $g^{IK}$
and $\frac{\partial g_{IK}}{\partial x^{L}},$ and for this reason
it is convenient to consider the Lagrangian as a function of $\phi
_{I},$ $g^{IK},$ and $\frac{\partial g_{IK}}{\partial x^{L}}$.

\subsection{\label{Energy-momentum tensor} Energy-momentum tensor}

Varying the Lagrangian $L_{d}$ $\left( \ref{Ldef}\right) $ with
respect to $\delta g^{IK}$ and having in mind that
 \begin{equation}
\delta g_{IK}=-g_{KM}g_{IN}\delta g^{NM},
\end{equation}
we get the following expression for the energy-momentum tensor
{\footnote{It differs from (94.4) in  \cite{Landau v2} because
the Lagrangian is considered there as a function of $g^{IK}$ and
$\frac{\partial g^{IK}}{\partial x^{L}}.$ Here and below  $\sqrt{-g}$ stands for  $\sqrt{(-1)^{D-1}g}.$ }}:
\begin{equation}
\begin{array}{c}
T_{IK}=\frac{2}{\sqrt{-g}}\left[ \frac{\partial
\sqrt{-g}L_{d}}{\partial g^{IK}}+g_{QK}g_{PI}\frac{\partial
}{\partial x^{L}}\left( \sqrt{-g}\frac{\partial L_{d}}{\partial
\frac{\partial g_{PQ}}{\partial x^{L}}}\right) \right]
 \label{Tik}
\end{array}
\end{equation}
In the case of the vector order parameter the symmetry breaking potential $V\left(
\phi ^{K}\phi _{K}\right) =V\left( g^{IK}\phi _{I}\phi _{K}\right)
$ also takes part in the variation with respect to $\delta g^{IK}$.

In what follows we apply this approach to the
 particular topological defect, namely the global string in two extra dimensions.

\section{\label{Global string in extra dimensions}  Global string in extra dimensions}

In my previous papers with Bronnikov (see \cite{Bron 1} and
references there in) we considered global monopoles and strings as
topological defects with the order parameter in the form of a
hedge-hock type multiplet of scalar fields in some flat target
space. The aim of this paper is to describe these defects using
vector order parameter and compare the results.

\subsection{\label{Metric} Metric}

The direction of the vector specifies one coordinate, and in the
most simple case the system is uniform and isotropic with respect
to all other coordinates. In our recent paper \cite{Bron 1} we
presented the detailed properties of global strings in two extra
dimensions. For this reason I consider below a topological defect
in the space-time with two extra dimensions. The order parameter
is a space-like vector $(g^{IK}\phi_{I}\phi_{K}<0)$ directed normally from the brane hypersurface and
depending on the only one specific coordinate, namely -- the
distance from the brane. The whole $\left( D=d_{0}+2\right)
$-dimensional space-time has the structure M$^{d_{0}}\times $
R$^{1}\times $ $\Phi ^{1}$ and the metric

\begin{equation}
ds^{2}
=e^{2\gamma \left( l\right) }\eta _{\mu
\nu }dx^{\mu }dx^{\nu }-\left( dl^{2}+e^{2\beta \left( l\right)
}d\varphi ^{2}\right) ,  \label{metric}
\end{equation}
where
$\eta_{\mu \nu }=$ diag $\left( 1,-1,...,-1\right)$ is the
$d_{0}$-dimensional Minkovsky brane metric $( d_{0}>1,$ we set $d_{0}=4$ in computations)
, and $\varphi $ is the angular cylindrical coordinate in extra
dimensions. $\gamma $ and $\beta $ are functions of the
distinguished extradimensional coordinate $l$ -- the distance from
the center, i.e. from the brane. $e^{\beta \left( l\right)
}=r\left( l\right) $ is the circular radius. Greek indices $\mu
,\nu ,..$ correspond to $d_{0}$-dimensional space-time on the
brane, and $I,K,...$ -- to all $D=d_{0}+2$ coordinates. The metric
tensor $g_{IK}$ is diagonal, and its nonzero components are
denoted as follows:

\begin{equation*}
g_{IK} =\left\{ \begin{array}{l} e^{2\gamma },\quad I=K=0, \\
-e^{2\gamma },\quad 0<I=K<d_{0}, \\  -1,\quad I=K=d_{0}, \\
-e^{2\beta },\quad I=K=\varphi.
\end{array} \right.
\end{equation*}

The curvature of the metric on brane due to the matter is supposed
to be much smaller than the curvature of the bulk caused by the
brane formation.

\subsection{\label{Regularity conditions} Regularity conditions}

If the influence of matter on brane is neglected, then there is no
physical reason for singularities, and the selfconsistent
structure of a topological defect should be regular. A necessary
condition of regularity is finiteness of all invariants of the
Riemann tensor of curvature. The nonzero components of the Riemann
tensor are

\begin{equation}\begin{array}{c}
R_{\text{ \ \ }CD}^{AB}= \left\{ \begin{array}{l} -\gamma ^{\prime
2}\left( \delta _{C}^{A}\delta _{D}^{B}-\delta _{D}^{A}\delta
_{C}^{B}\right),\quad A,B,C,D<d_{0}, \\ -\beta ^{\prime }\gamma
^{\prime },\quad A=C=\varphi ,\quad B,D<d_{0}, \\ -\left( \gamma
^{\prime \prime }+\gamma ^{\prime 2}\right) \delta _{D}^{B},\quad
A=C=d_{0},\quad B,D<d_{0}, \\ -\left( \beta ^{\prime \prime
}+\beta ^{\prime 2}\right) ,\quad A=C=d_{0},\quad B=D=\varphi .
\end{array}\right.
\label{R^AB_CD}
\end{array}
\end{equation}

Here prime denotes $d/dl.$ One of the invariants of the Riemann
tensor is the Kretchmann scalar $K=R_{\text{ \ \
}CD}^{AB}R_{\text{ \ \ }AB}^{CD}$ , which is the sum of all
nonzero $R_{\text{ \ \ }CD}^{AB}$ squared. Hence all the nonzero
components of the Riemann tensor, and namely
\begin{equation}
\gamma ^{\prime },\ \ \gamma^{\prime \prime }+\gamma^{\prime 2}, \
\ \beta^{\prime }\gamma^{\prime }, \ \ \beta^{\prime \prime
}+\beta^{\prime 2}
 \label{Regul conditions}
\end{equation}
must be finite. The center $r=0$ is a singular point of the cylindrical
coordinate system. The absence of curvature singularity in the
center follows from the finiteness of the last term in $\left( \ref{Regul
conditions}\right) .$ Let
\begin{equation} \beta ^{\prime \prime }+\beta ^{\prime
2}=c<\infty \text{\ \ \ at \ \ }l=0. \label{Bet''+Bet'^2=c}
\end{equation}
Integrating $\left( \ref{Bet''+Bet'^2=c}\right) $ in the vicinity
of the center we have \begin{equation} \beta ^{\prime
}=\frac{1}{l}+\frac{1}{3}cl+O\left( l^{3}\right) .
\label{Bet'=1/l+..}
\end{equation}
Relation $\left( \ref{Bet'=1/l+..}\right) $ ensures the correct
$\left( =2\pi \right) $ circumference-to-radius ratio, or,
equivalently, $dr^{2}=dl^{2}$ at $l\rightarrow 0.$ Finiteness of
$\beta ^{\prime }\gamma ^{\prime }$ at $l=0$ is fulfilled if
\begin{equation} \gamma ^{\prime }=O\left( l\right) \text{\ \ \ or
\ \ smaller \ \ at \ \ }l\rightarrow 0. \label{Gamma'=O(l)}
\end{equation}

\subsection{\label{Vector order parameter} Vector order parameter}

Our aim is to consider the order parameter as a vector in extra
dimensions directed normally from the Minkovsky hypersurface. In
the cylindrical coordinate system of extra dimensions the only
nonzero component of the vector order parameter is\
\begin{equation} \phi _{d_{0}}\equiv \phi .  \label{Fi_d0=Fi}
\end{equation}
In the space-time with the metric $\left( \ref{metric}\right) $\
the covariant derivative \begin{equation} \phi _{I;K}=\delta
_{I}^{d_{0}}\delta _{K}^{d_{0}}\phi ^{\prime }-\frac{1}{2}\delta
_{IK}g^{II}g_{II}^{\prime }\phi  \label{Fi_I;K=}
\end{equation}
is a symmetric tensor: $\phi _{I\text{ };K}=\phi _{K\text{ };I}.$
For this reason $\phi _{\text{ };K}^{I}\phi _{I\text{ }}^{\text{
};K}=\phi _{\text{ };K}^{I}\phi _{\text{ };I}^{K}$ , and without loss of generality we set $C=0.$ The Lagrangian $\left( \ref{Ldef}\right) $ takes the form
\begin{equation}
\begin{array}{c}L_{d}=A\left( \phi ^{\prime }+\frac{1}{2}\phi
\sum_{K}g^{KK}g_{KK}^{\prime }\right) ^{2} +B\left(
\phi ^{\prime 2}+\frac{1}{4}\phi ^{2}\sum_{L}\left(
g^{LL}g_{LL}^{\prime }\right) ^{2}\right) -V\left( -\phi
^{2}\right). \end{array} \label{L_d=...}
\end{equation}
It contains only two arbitrary constants $A$ and
$B.$ In $\left( \ref{L_d=...}\right) $ we set
$g^{d_{0}d_{0}}=-1$ in accordance with $\left( \ref{metric}\right)
.$ However one should keep in mind that $\left(
\ref{L_d=...}\right) $ cannot be used in $\left( \ref{Tik}\right)
.$ To derive the energy-momentum tensor $\left( \ref{Tik}\right)
$\ one should use the Lagrangian $\left( \ref{Ldef}\right) ,$ and
set $g^{d_{0}d_{0}}=-1,$ $\left( g^{d_{0}d_{0}}\right) ^{\prime
}=0$\ after differentiation. Nevertheless, the field equation can
be derived using $\left( \ref{L_d=...}\right) $ in the general
formula \begin{equation} \frac{1}{\sqrt{-g}}\left( \frac{\partial
\sqrt{-g}L_{d}}{\partial \phi ^{\prime }}\right) ^{\prime
}-\frac{\partial L_{d}}{\partial \phi }=0.
\label{(dL_d/dFi')'-dL_d/dFi=0}
\end{equation}
In the space-time with metric $\left( \ref{metric}\right) $ the
sums in $\left( \ref{L_d=...}\right) $ are
\begin{equation}\begin{array}{c}
S_{n}=\frac{1}{2^{n}}\sum_{K}\left( g^{KK}g_{KK}^{\prime }\right)
^{n}= d_{0}\gamma ^{\prime n}+\beta ^{\prime n},\quad
n=1,2,...\text{ .} \label{Sums}
\end{array}\end{equation}
and the determinant of the metric tensor is \begin{equation}
g=\left( -1\right) ^{D-1}e^{2\left( d_{0}\gamma +\beta \right) }.
\label{g=e^...}
\end{equation}
We consider below the two cases, case \textbf{A} ($A\neq 0,$ $B=0$) and case \textbf{B} ($B\neq 0,$ $A=0$), in comparison, but separately.

\subsection{\label{Field equations} Field equations}

Substituting $\left( \ref {L_d=...}\right) $  into $\left( \ref
{(dL_d/dFi')'-dL_d/dFi=0}\right) $ we get the field
equations in the case of vector order parameter.

\textbf{\textit{Case} A}  \\
In case $A=\frac{1}{2},$
$B=0$  we get
\begin{equation}
\left( \phi ^{\prime }+S_{1} \phi \right) ^{\prime }+\frac{\partial
V}{\partial \phi }=0. \label{Field equationA}
\end{equation}

\textbf{\textit{Case} B}  \\
In the case $B=\frac{1}{2},$
$A=0$ the substitution results in
\begin{equation}
\phi ^{\prime \prime }+S_{1} \phi ^{\prime }-S_{2} \phi +\frac{\partial V}{\partial \phi }=0. \label{Field equationB}
\end{equation}

$S_{1}$ and $S_{2}$ are defined in (\ref{Sums}).

\textbf{\textit{Case of scalar multiplet} }  \\
In the case of the multiplet of scalar fields we had \cite{Bron 1}:
\begin{equation} \phi ^{\prime \prime }+S_{1} \phi ^{\prime } -\phi e^{-2\beta
}+\frac{\partial V}{\partial \phi }=0. \label{Scalar field eq}
\end{equation}

Unlike (\ref{Scalar field eq}), the field equations (\ref{Field
equationA}) and (\ref{Field
equationB})  do not depend directly on $\beta$ (and thus on the
circular radius $r=\ln \beta )$. The case A field equation (\ref{Field equationA})  includes second
derivatives of the metric tensor. In the flat space-time $\gamma
^{\prime }=0,$ $\beta ^{\prime }=\frac{1}{l},$ $\beta ^{\prime
\prime }=-\frac{1}{l^{2}},e^{-2\beta }=\frac{1}{l^{2}},$ \ and all three
 field equations (\ref{Field
equationA}),  (\ref{Field
equationB}), and (\ref{Scalar field eq}) reduce to
\begin{equation} \phi ^{\prime \prime }+\frac{1}{l}\phi ^{\prime
}-\frac{1}{l^{2}}\phi +\frac{\partial V}{\partial \phi }=0.
\label{Flat field equation}
\end{equation}

\subsection{\label{sub:Energy-momentum tensor} Energy-momentum tensor}

The energy-momentum tensor $\left( \ref{Tik}\right) $ inevitably
contains second derivatives. However, in case A the second
derivatives can be excluded with the aid of the field
equation $\left( \ref {Field equationA}\right)$. In case B the second derivatives cannot be excluded from the energy-momentum tensor. The final result of  rather wearing
derivations is:

\textbf{\textit{Case} A}
\begin{equation}
 \begin{array}{c}T_{I}^{K}=\frac{1}{2}\delta
_{I}^{K}\left( \phi ^{\prime }+S_{1}\phi \right) ^{2} +\delta _{I}^{K}V+\left(
\delta _{I}^{d_{0}}\delta _{d_{0}}^{K}-\delta _{I}^{K}\right)
\frac{\partial V}{\partial \phi }\phi
 \end{array}\label{T ik=1/2Delta^K_I...A}
\end{equation}

\textbf{\textit{Case} B}
\begin{equation}
\begin{array}{l}
T_{I<d_{0}}^{K} = \delta _{I}^{K}\left[ \frac{1}{\sqrt{-g}}\left( \sqrt{-g}\gamma ^{\prime }\right) ^{\prime }\phi ^{2}+\gamma ^{\prime }\left( \phi ^{2}\right) ^{\prime }-\left( \frac{1}{2}\phi ^{\prime 2}+\frac{1}{2}S_{2}\phi ^{2}\right) +V\right]  \\
T_{d_{0}}^{d_{0}} =\frac{1}{2}\left( \phi ^{\prime 2}+S_{2}\phi ^{2}\right) +V  \\
T_{I>d_{0}}^{K} =\delta _{I}^{K}\left[ \frac{1}{\sqrt{-g}}\left( \sqrt{-g}\beta ^{\prime }\right) ^{\prime }\phi ^{2}+\beta ^{\prime }\left( \phi ^{2}\right) ^{\prime }-\left( \frac{1}{2}\phi ^{\prime 2}+\frac{1}{2}S_{2}\phi ^{2}\right) +V\right]
\end{array}
\label{T ik=1/2Delta^K_I...B}
\end{equation}

Unlike the scalar multiplet case, the energy-momentum tensor
$\left( \ref{T ik=1/2Delta^K_I...A}\right) $ contains not only the
potential $V,$ but also its derivative $\frac{\partial V}{\partial
\phi }$.

Correctness of $\left( \ref{T ik=1/2Delta^K_I...A}\right) $ and $\left( \ref{T ik=1/2Delta^K_I...B}\right)$ is
checked by the derivation of the covariant divergence $T_{I\text{
};K}^{K}$ (actually $T_{d_{0}\text{ };K}^{K})$. Again, with the
aid of field equations $\left( \ref{Field equationA}\right) $ and $\left( \ref{Field equationB}\right) $ we
confirm that \begin{equation*} T_{d_{0}\text{ };K}^{K}=0.
\end{equation*}

\subsection{\label{Einstein equations} Einstein equations}

The same way as in \cite{Bron 1} we use the Einstein equations
in the form
\begin{equation*} R_{I}^{K}=\kappa
^{2}\widetilde{T}_{I}^{K},
\end{equation*}
where $R_{I}^{K}$ is the Ricci tensor,
\begin{equation*}
R_{I}^{K}=\left \{ \begin{array}{c}
\delta _{I}^{K}\left( \gamma
^{\prime \prime }+\gamma ^{\prime }S_{1} \right) ,\quad I<d_{0} \\
\delta_{d_{0}}^{K}\left( S_{1}^{\prime}+S_{2}\right) ,\quad I=d_{0} \\ \delta _{\varphi }^{K}\left( \beta
^{\prime \prime }+S_{1}\beta ^{\prime } \right) ,\quad I=\varphi
\end{array} \right.
 \end{equation*}
and
\begin{eqnarray*} \begin{array}{c}
\widetilde{T}_{I}^{K}=T_{I}^{K}-\frac{1}{d_{0}}\delta _{I}^{K}T
 . \end{array}
\end{eqnarray*}
In both cases A and B the sets of Einstein equations
consist of three first order equations with respect to $\gamma
^{\prime },$ $\beta ^{\prime },$ and $\phi $  \footnote{The equations do not contain the coordinate $l$ directly. The substitution $p=\frac{d\phi}{dl},$ $\frac{d}{dl}=p\frac{d}{d\varphi}$ reduces  the Einstein equations in both cases A (\ref{Gamma''+...A}-\ref{Beta''+...A}) and B (\ref{Gamma''+...B}-\ref{Beta''+...B}) to the sets of the second order for unknowns $\gamma^{\prime}(\phi)$ and $\beta^{\prime}(\phi).$ It simplifies to prove that the field equations are the consequences of the Einstein equations, but it does not really  help neither in analytical, nor in numerical analysis of the equations.}:

\textbf{\textit{Case} A}
\begin{eqnarray}
\begin{array}{c}
\gamma ^{\prime \prime }+S_{1}\gamma ^{\prime } = \kappa ^{2}\left[
-\frac{1}{d_{0}}\left( \phi ^{\prime }+ S_{1} \phi \right)
^{2}-\frac{2V}{d_{0}}+\frac{1}{d_{0}}\frac{\partial V}{\partial
\phi }\phi \right] \end{array} \label{Gamma''+...A} \\
\begin{array}{c} S_{1}^{\prime}+S_{2}=  \kappa ^{2}\left[ -\frac{1}{d_{0}}\left( \phi
^{\prime }+S_{1}
\phi \right) ^{2}-\frac{2V}{d_{0}}+\left( 1+\frac{1}{d_{0}}\right)
\frac{\partial V}{\partial \phi }\phi \right]
\end{array}  \\
\begin{array}{c}\beta ^{\prime \prime }+ S_{1}\beta
^{\prime } =
 \kappa ^{2}\left[ -\frac{1}{d_{0}}\left( \phi ^{\prime
}+S_{1} \phi
\right) ^{2}-\frac{2V}{d_{0}}+\frac{1}{d_{0}}\frac{\partial
V}{\partial \phi }\phi \right] \end{array} \label{Beta''+...A}
\end{eqnarray}

\textbf{\textit{Case} B}
\begin{eqnarray}
\begin{array}{c}
\gamma ^{\prime \prime }+S_{1}\gamma ^{\prime }=\kappa ^{2}\left[ S_{1}\gamma ^{\prime }\phi ^{2}+\left( \gamma ^{\prime }\phi ^{2}\right) ^{\prime }-\frac{1}{d_{0}}S_{1}^{2}\phi ^{2}-\frac{1}{d_{0}}\left( S_{1}\phi ^{2}\right) ^{\prime }-\frac{2}{d_{0}}V\right] \end{array} \label{Gamma''+...B} \\
\begin{array}{c} S_{1}^{\prime}+S_{2}= \kappa ^{2}\left[ \phi ^{\prime 2}+S_{2}\phi ^{2}-\frac{1}{d_{0}}S_{1}^{2}\phi ^{2}-\frac{1}{d_{0}}\left( S_{1}\phi ^{2}\right) ^{\prime }-\frac{2}{d_{0}}V\right]  \end{array}  \\
\begin{array}{c}\beta ^{\prime \prime }+ S_{1}\beta
^{\prime } =\kappa ^{2}\left[ S_{1}\beta ^{\prime }\phi ^{2}+\left( \beta ^{\prime }\phi ^{2}\right) ^{\prime }-\frac{1}{d_{0}}S_{1}^{2}\phi ^{2}-\frac{1}{d_{0}}\left( S_{1}\phi ^{2}\right) ^{\prime }-\frac{2}{d_{0}}V\right]
\end{array} \label{Beta''+...B}
\end{eqnarray}

Metric functions $\gamma $ and
$\beta $ do not enter the equations $\left(
\ref{Gamma''+...A}-\ref{Beta''+...B}\right) $\ directly, only via
the derivatives. In the case of a scalar multiplet order
parameter, see eq.$\left( 14-16\right) $ in \cite{Bron 1},
$\beta $ enters the Einstein equations directly, and the system of
equations is of the fourth order. It is also worth to mention that the right hand sides of first and third equations coincide in case A, and have a similar structure in case B. This symmetry reflects the fact that the coordinates $x^{I}$ with $I<d_{0}$ and $I=\varphi$ are cyclic variables.

The field equations $\left( \ref{Field equationA}\right) $ and $\left( \ref{Field equationB}\right) $ are not
independent. They are consequences of the Einstein equations due to the Bianchi identities.

\subsection{\label{First integrals} First integrals}

Excluding the second derivatives $\gamma ^{\prime \prime }$ and
$\beta ^{\prime \prime }$ in the sets $\left(
\ref{Gamma''+...A}-\ref{Beta''+...A}\right)$ and $\left(
\ref{Gamma''+...B}-\ref{Beta''+...B}\right) ,$ we get the relations

\textbf{\textit{Case} A}
\begin{equation}
\begin{array}{c}S_{1}^{2}-S_{2} =  -\kappa ^{2}\left[ \left( \phi ^{\prime }+S_{1} \phi \right)
^{2}+2V\right] , \end{array}  \label{First integralA}
\end{equation}

 \textbf{\textit{Case} B}
\begin{equation}
\begin{array}{c}
S_{2}\left( 1-\kappa ^{2}\phi ^{2}\right) =\kappa ^{2}\phi ^{\prime 2}+2\kappa ^{2}V+S_{1}^{2},
\end{array}  \label{First integralB}
\end{equation}
which can be considered as  first integrals of the systems $\left(
\ref {Gamma''+...A}-\ref{Beta''+...A}\right) $ and $\left(
\ref {Gamma''+...B}-\ref{Beta''+...B}\right) .$

\subsection{\label{Further simplification} Further simplification}

The equations $\left( \ref{Gamma''+...A}\right) $ and $\left(
\ref {Beta''+...A}\right), $ as well as the equations $\left( \ref{Gamma''+...B}\right) $ and $\left(
\ref {Beta''+...B}\right),$ have similar structures.
Extracting in both cases  one from the other, we get the equations

 \textbf{\textit{Case} A}
\begin{equation} \left( \gamma ^{\prime }-\beta ^{\prime }\right)
^{\prime }+\left( \gamma ^{\prime }-\beta ^{\prime }\right) S_{1} =0,
\label{(Gamma'-Beta')'+...A}
\end{equation}

\textbf{\textit{Case} B}
\begin{equation} \left[ \left( \gamma ^{\prime }-\beta ^{\prime }\right) \left( 1-\kappa ^{2}\phi ^{2}\right) \right] ^{\prime }+\left( \gamma ^{\prime }-\beta ^{\prime }\right) \left( 1-\kappa ^{2}\phi ^{2}\right) S_{1}=0.
\label{(Gamma'-Beta')'+...B}
\end{equation}
$\left(
\ref{(Gamma'-Beta')'+...A}\right) $  can be used instead of one of the equations $\left(
\ref{Gamma''+...A}\right) $ or $\left( \ref{Beta''+...A}\right) $  and, respectively, $\left(
\ref{(Gamma'-Beta')'+...B}\right) $ -- instead of  $\left(
\ref{Gamma''+...B}\right) $ or $\left( \ref{Beta''+...B}\right) $ .

With the aid of the relations $\left( \ref{First integralA}\right)
$,$\left( \ref {(Gamma'-Beta')'+...A}\right) $ and $\left( \ref{First integralB}\right)
$,$\left( \ref {(Gamma'-Beta')'+...B}\right) $  the complete sets
of equations can be reduced to more simple forms. Introducing new functions
\begin{equation} U=\gamma ^{\prime }-\beta ^{\prime
}, \text{ \ }Z=\phi ^{\prime }+S_{1}\phi , \text{ \ } \psi=\phi^{\prime }, \label{new functions}
\end{equation}
and having in mind that functions $\beta ^{\prime },$ $\gamma ^{\prime },$ and their
combination $S_{2}=d_{0}\gamma ^{\prime 2}+\beta ^{\prime 2}$
$\left( \ref{Sums}\right) $\ are expressed via $U$ and $S_{1}$ as
follows:    \\
\begin{equation*} \gamma ^{\prime
}=\frac{U+S_{1}}{d_{0}+1}\qquad \beta ^{\prime
}=\frac{S_{1}-d_{0}U}{d_{0}+1}\qquad
S_{2}=\frac{d_{0}U^{2}+S_{1}^{2}}{d_{0}+1},
\end{equation*} \\
we get the sets of four first order equations, solved against the derivatives:

\textbf{\textit{Case} A}
\begin{equation}
\begin{array}{c}
U^{\prime }=-U S_{1} \\ S_{1}^{\prime }=\kappa
^{2}\frac{d_{0}+1}{d_{0}}\frac{\partial V}{\partial \phi }\phi
-U^{2} \\ \phi ^{\prime }=Z-S_{1}\phi \\ Z^{\prime }=-\frac{\partial
V}{\partial \phi }
\end{array}
\label{Convenient setA}
\end{equation}

\textbf{\textit{Case} B}
\begin{equation}
\begin{array}{c}
\left[ U\left( 1-\kappa ^{2}\phi ^{2}\right) \right] ^{\prime }+U\left( 1-\kappa ^{2}\phi ^{2}\right) S_{1}=0 \\
\left[ S_{1}\left( 1+\frac{\kappa ^{2}\phi ^{2}}{d_{0}}\right) \right] ^{\prime }+S_{1}^{2}\left( 1+\frac{\kappa ^{2}\phi ^{2}}{d_{0}}\right) +\frac{2\left( 1+d_{0}\right) }{d_{0}}\kappa ^{2}V=0 \\
\phi ^{\prime }=\psi  \\
\psi^{\prime }+S_{1}\psi -\frac{d_{0}U^{2}+S_{1}^{2}}{d_{0}+1}\phi +\frac{\partial V}{\partial \phi }=0
\end{array}
\label{Convenient setB}
\end{equation}

The sets $\left( \ref{Convenient setA}\right) $
and $\left( \ref{Convenient setB}\right) $ are most convenient for both analytical and numerical analysis.

\subsection{\label{General analysis of regular solutions} General analysis of regular solutions}

The sets $\left( \ref{Convenient setA}\right) $
and $\left( \ref{Convenient setB}\right) $ are
invariant against adding arbitrary constants to $\gamma $ and
$\beta .$ Without loss of generality we can set \begin{equation}
\gamma \left( 0\right) =0.  \label{Gamma(0)=0}
\end{equation}
Requirement of regularity in the center dictates the condition
$\left( \ref {Bet'=1/l+..}\right) ,$ and, if we do not consider
configurations with angle deficit (or surplus), we have
\begin{equation} r=e^{\beta }=l\ \text{\ at }\ l\rightarrow 0.
\label{r=l at l to 0}
\end{equation}
Integrating first equations in the sets  $\left( \ref{Convenient setA}\right) $
and $\left( \ref{Convenient setB}\right) $ with
boundary conditions $\left( \ref{Gamma(0)=0},\ref{r=l at l to
0}\right) $ we get:

\textbf{\textit{Case} A}
\begin{equation} \gamma ^{\prime }-\beta
^{\prime }=-e^{-\left( d_{0}\gamma +\beta \right) },
\label{G'-B'=-e^(d0G+B)A}
\end{equation}

\textbf{\textit{Case} B}
\begin{equation} \gamma ^{\prime }-\beta
^{\prime }=-\frac{e^{-\left( d_{0}\gamma +\beta \right)}}{1-\kappa^{2}\phi^{2}} .
\label{G'-B'=-e^(d0G+B)B}
\end{equation}
From $\left( \ref{G'-B'=-e^(d0G+B)A}\right) $ and $\left( \ref{G'-B'=-e^(d0G+B)B}\right) $ it follows that in  both cases
$\beta ^{\prime }>\gamma ^{\prime }$ everywhere. Within the regions of regularity $e^{-\left( d_{0}\gamma +\beta \right)}\rightarrow0$ at $l\rightarrow\infty,$ and $\gamma_{\infty} ^{\prime }$ coincides with $\beta_{\infty}
^{\prime }.$

The denominator in $\left( \ref{G'-B'=-e^(d0G+B)B}\right) $ is the origin of an important difference between the cases A and B. In the symmetric (unbroken) state the order parameter $\phi=0,$ and $\phi^{\prime}=0$ everywhere. In the state with broken symmetry $\phi^{\prime}\neq 0$, but still  $\phi=0$ at a point of maximum of the symmetry breaking potential $V(\phi)$, which is supposed to coincide with the center $l=0$. Requirement of regularity is fulfilled only if the denominator in $\left( \ref{G'-B'=-e^(d0G+B)B}\right) $ remains positive everywhere. It means that in case B the order parameter $\phi(l)$ can never exceed $\kappa^{-1}:$
\begin{equation}
\kappa ^{2}\phi ^{2}<1,\qquad \textrm{case B.}
\label{kappa_fi<1}
\end{equation}
Case A is free from such a restriction.

Recall that topological defects, formed as multiplets of scalar
fields \cite {Bron 1}, are of three types. Integral curves can
terminate with:

A) infinite circular radius $r\left( l\right) $ at $l\rightarrow
\infty ;$

B) finite circular radius $r_{\infty }=r\left( \infty \right)
=const<\infty ; $

C) second center $r=0$ at some finite $l=l_{c}.$

In the vector order parameter case the situation is different.
Equations $\left( \ref{G'-B'=-e^(d0G+B)A}\right) $ and $\left( \ref{G'-B'=-e^(d0G+B)B}\right) $ allow to prove
that  regular configurations in both cases A and B can not terminate neither with a
finite value of circular radius $r_{\infty }$\ at $l\rightarrow
\infty ,$ nor in the second center$.$

Suppose, for instance in case A, that $r_{\infty }=const<\infty .$ Then
$\beta ^{\prime }(\infty )=0,$ and $\left(
\ref{G'-B'=-e^(d0G+B)A}\right) $ reduces to $\gamma ^{\prime
}=-\frac{1}{r_{\infty }}e^{-d_{0}\gamma }$\ at $l\rightarrow
\infty .$ After integration we get \begin{equation*}
e^{d_{0}\gamma }=\frac{d_{0}}{r_{\infty }}\left( l_{0}-l\right) ,
\end{equation*}
$l_{0}$ is a constant of integration. The l.h.s. is obviously
positive, while the r.h.s. becomes negative and infinitely large
at $l\rightarrow \infty .$ Thus $r_{\infty }=const<\infty $ is
impossible.

The second center is also impossible. In the vicinity of the
second center the l.h.s. of $\left( \ref{G'-B'=-e^(d0G+B)A}\right)
$ becomes large positive due to $-\beta ^{\prime },\ $and the
r.h.s. remains negative.

We come to the conclusion that regular configurations of
topological defects with the vector order parameter start at the
center $l=0$ and always  terminate at
 $l\rightarrow \infty $ with infinitely growing circular radius $r\left(
l\right) \rightarrow \infty .$

It follows from the requirement of regularity $\left(
\ref{Gamma'=O(l)}\right) $ that $\gamma ^{\prime }$ $=\gamma
_{0}^{\prime \prime }l$ at $l\rightarrow 0.$ From the first
integrals $\left( \ref{First integralA}\right) $ and $\left( \ref{First integralB}\right) $  we find the
relations between $\gamma _{0}^{\prime \prime },\phi _{0}^{\prime
},$ and $V_{0}:$

\textbf{\textit{Case} A}
\begin{equation}
\gamma _{0}^{\prime \prime }=-\frac{\kappa ^{2}}{d_{0}}\left(
2\phi _{0}^{\prime 2}+V_{0}\right) ,  \label{relation for bound
condA}
\end{equation}

\textbf{\textit{Case} B}
\begin{equation}
\gamma _{0}^{\prime \prime }=-\frac{\kappa ^{2}}{d_{0}}\left( \frac{1}{2}\phi _{0}^{\prime 2}+V_{0}\right)  ,  \label{relation for bound
condB}
\end{equation}
where $V_{0}$ is the value of the symmetry breaking potential at the center $l=0.$ In both cases A and B, as well as in the case of scalar multiplet,  the value
$\phi _{0}^{\prime }=\phi ^{\prime }\left( 0\right) $ is not
restricted by the equations. The difference is that in the scalar
multiplet case $\phi _{0}^{\prime }$ becomes fixed unanimously by
the requirement of regularity, while in  both cases of vector order
parameter \bigskip $\phi _{0}^{\prime }$ remains a free parameter.

\subsection{\label{Asymptotic behavior} Asymptotic behavior}

Condition of regularity requires that $\gamma ^{\prime }$ is
finite everywhere. Within the area of regularity it tends to a
fixed finite value $\gamma _{\infty }^{\prime }$ at $l\rightarrow
\infty $. As soon as $r\left( l\right) \rightarrow \infty $ at
$l\rightarrow \infty ,$ we see from $\left(
\ref{G'-B'=-e^(d0G+B)A}\right) $ and $\left(
\ref{G'-B'=-e^(d0G+B)B}\right) $ that $\gamma ^{\prime }-\beta
^{\prime }\rightarrow 0$ in both cases A and B. Thus
 $\beta ^{\prime }\left( \infty
\right) =\gamma_{\infty }^{\prime }.$
 The field
$\phi \left( l\right) $ also tends to its finite value $\phi
_{\infty }=\phi \left( \infty \right) .$

\textbf{\textit{Case} A}  \\
From the
field equation $\left( \ref{Field equationA}\right) $ it follows  that
$\frac{\partial V}{\partial \phi }\rightarrow 0$ at $l\rightarrow
\infty ,$ i.e. in the case A   regular configurations terminate at an extremum
of the potential $V\left( \phi \right) $. Let $V_{\infty }=V\left(
\phi _{\infty }\right) ,$ $V^{\prime }\left( \phi _{\infty
}\right) =0.$ From the first integral $\left( \ref{First
integralA}\right) $ we find the limiting value $\gamma _{\infty
}^{\prime }:$
\begin{equation}
\gamma _{\infty }^{\prime }=\sqrt{-\frac{2\kappa ^{2}V_{\infty
}}{\left( d_{0}+1\right) \left[ d_{0}+\left( d_{0}+1\right) \kappa
^{2}\phi _{\infty }^{2}\right] }}.  \label{Gam'_inf =A}
\end{equation}

A necessary condition of existence of regular configurations of
topological defects with the vector order parameter is $V_{\infty
}<0.$

\textbf{\textit{Case} B}  \\
At $l\rightarrow \infty $  the second  equation in $\left(\ref{Convenient setB}\right) $ reduces to
\begin{equation*}
S_{1\infty }^{2}\left( 1+\frac{\kappa ^{2}\phi _{\infty }^{2}}{d_{0}}\right) +\frac{2\left( 1+d_{0}\right) }{d_{0}}\kappa ^{2}V_{\infty }=0.
\end{equation*}
We see that in this case also
\begin{equation}
V_{\infty }=V\left( \phi _{\infty }\right) <0.
\label{V<0}
\end{equation}

The fourth equation in $\left(\ref{Convenient setB}\right) $ reduces to
\begin{equation*}
-S_{1\infty }^{2}\frac{\phi _{\infty }}{d_{0}+1}+\frac{\partial V_{\infty }}{\partial \phi }=0,
\end{equation*}
and so
\begin{equation}
\frac{1}{2\phi _{\infty }}\frac{\partial V_{\infty }}{\partial \phi }\equiv \frac{\partial V_{\infty }}{\partial \phi ^{2}}>0.
\label{V'>0}
\end{equation}
Here $V_{\infty }$ and$\ \frac{\partial V_{\infty }}{\partial \phi }$ are the  values of the symmetry breaking potential and its derivative at $l\rightarrow \infty .$ Unlike the case A, in case B regular configurations terminate at a  slope of the potential, and not at a point of extremum. Excluding $S_{1}^{2}$ we get the equation determining $\phi _{\infty }=\phi \left( \infty \right) :$
\begin{equation}
\kappa ^{2}V_{\infty }=-\left( d_{0}+\kappa ^{2}\phi _{\infty }^{2}\right) \frac{\partial V_{\infty }}{\partial \phi ^{2}}. \label{eq_for_fi_infB}
\end{equation}
In view of $\ S_{1\infty }=\left( d_{0}+1\right) \gamma _{\infty }^{\prime },$ we find
\begin{equation}
\gamma _{\infty }^{\prime }=\sqrt{\frac{1}{\left( d_{0}+1\right) \phi _{\infty }}\frac{\partial V_{\infty }}{\partial \phi }}.
\label{eq_for_gamma_infB}
\end{equation}

To find the  behavior of $\phi \left( l\right) $ and
$S_{1}\left( l\right) $ at large distances from the center we have to linearize the sets of equations $\left(
\ref{Convenient setA}\right) $ and   $\left(
\ref{Convenient setB}\right) $ at $l\rightarrow \infty :$
\begin{equation*}
\phi =\phi _{\infty }+\delta \phi ,\text{ \ }S_{1}=\left(
d_{0}+1\right) \gamma _{\infty }^{\prime }+\delta S_{1}.
\end{equation*}

\textbf{\textit{Case} A}:
\begin{equation}
\begin{array}{c}
\delta S_{1}^{\prime }=\kappa ^{2}\frac{d_{0}+1}{d_{0}}V_{\infty
}^{\prime \prime }\phi _{\infty }\delta \phi \\ \delta \phi
^{\prime }=\delta Z-\left( d_{0}+1\right) \gamma _{\infty
}^{\prime }\delta \phi -\phi _{\infty }\delta S_{1} \\ \delta
Z^{\prime }=-V_{\infty }^{\prime \prime }\delta \phi
\end{array}
\label{Linear set}
\end{equation}
Here primes denote derivatives $d/dl,$ $\left( \delta S_{1}^{\prime
}=d (\delta S_{1})/dl,...\right) ,$ except $V_{\infty }^{\prime \prime
}=\frac{\partial ^{2}V}{\partial \phi ^{2}}\left| _{\phi =\phi
_{\infty }}\right. .$ \ Excluding $\delta Z$ and $\delta S_{1},$ we
get the second order linear homogeneous equation for $\delta \phi:$

\begin{equation*}
\delta \phi ^{\prime \prime }+\left( d_{0}+1\right) \gamma
_{\infty }^{\prime }\delta \phi ^{\prime }+\frac{2\kappa
^{2}\left| V_{\infty }\right| V_{\infty }^{\prime \prime
}}{d_{0}\left( d_{0}+1\right) \gamma _{\infty }^{\prime 2}}\delta
\phi =0.
\end{equation*}

In case A the solution terminates at an extremum of $V(\phi)$. If the extremum of the potential is minimum, $\left( V_{\infty
}^{\prime \prime }>0\right), $\ the nontrivial solution vanishes at
$l\rightarrow \infty :$
\begin{equation}
 \delta \phi =Ae^{\lambda
_{+}l}+Be^{\lambda _{-}l}, \label{Solutions at l to infA}
\end{equation}
where $A$ and $B$ are constants of integration, and both
eigenvalues \begin{equation} \lambda _{\pm }=-\frac{\left(
d_{0}+1\right) \gamma _{\infty }^{\prime }}{2}\left( 1\mp
\sqrt{1-\frac{8\kappa ^{2}\left| V_{\infty }\right| V_{\infty
}^{\prime \prime }}{d_{0}\left( d_{0}+1\right) ^{3}\gamma _{\infty
}^{\prime 4}}}\right)  \label{eigenvaluesA}
\end{equation}
are either negative, or have negative real parts. Absence of
growing solutions is the reason why $\phi _{0}^{\prime }$ remains
a free parameter in the vector order parameter case.

In case A the asymptotic behavior of the field $\phi \left( l\right) $ far
from the center is determined by two constant parameters of the
symmetry breaking potential near its extremum, namely $V_{\infty
}$ and $V_{\infty }^{\prime \prime }.$ If the extremum is minimum,
$V_{\infty }^{\prime \prime }>0$, then the expression under the
root can be both positive and negative. So $\phi \left( l\right) $
can tend to $\phi _{\infty }$ either smoothly, or with
oscillations. In the space of physical parameters the boundary
between smooth and oscillating solutions is determined by the
relation \begin{equation} \frac{8\kappa ^{2}\left| V_{\infty
}\right| V_{\infty }^{\prime \prime }}{d_{0}\left( d_{0}+1\right)
^{3}\gamma _{\infty }^{\prime 4}}=1. \label{Boundary between mon
and osc}
\end{equation}

Oscillating behavior of the field $\phi \left( l\right) $ induces
oscillations of $\beta ^{\prime }$ and $\gamma ^{\prime }.$ If
$\gamma ^{\prime }$ changes sign, then $\gamma \left( l\right) $
can have minimums. Remind, that in the weak gravitation limit $\gamma $ acts as a gravitational
potential, so the matter can be trapped near the minimums of
$\gamma \left( l\right) $. Trapping of matter to the brane is considered below in sections \ref{Neutral quantum particle} and \ref{Oscillations} in more detail.

Usually $\phi =0$ is a maximum of the potential $V\left( \phi
\right) .$ It is also an extremum, $\partial V/\partial \phi =0$
at $\phi =0$. Regular configurations, starting from the center
$l=0$ with $\phi \left( 0\right) =0, $ can terminate at
$l\rightarrow \infty $ with $\phi _{\infty }=0$ as well. In this
case $V_{\infty }^{\prime \prime }=V^{\prime \prime }\left(
0\right) <0,$ and the linear set $\left( \ref{Linear set}\right) $
reduces to the following asymptotic equation for $\phi \left(
l\right) :$
\begin{equation*} \phi ^{\prime \prime }+\left(
d_{0}+1\right) \gamma _{\infty }^{\prime }\phi ^{\prime }-\left|
V_{\infty }^{\prime \prime }\right| \phi =0.
\end{equation*}
Its general solution is a linear combination of vanishing and
growing functions:
\begin{equation*}
 \phi =Ae^{-\lambda
_{+}l}+Be^{-\lambda _{-}l}, \\
\qquad \lambda _{\pm }=\frac{\left(
d_{0}+1\right) \gamma _{\infty }^{\prime }}{2}\pm
\sqrt{\frac{\left( d_{0}+1\right) ^{2}\gamma _{\infty }^{\prime
2}}{4}+\left| V_{\infty }^{\prime \prime }\right| },\quad \quad
l\rightarrow \infty .
\end{equation*}
Requirement of regularity demands to exclude the growing solutions
from the consideration. It can be done at the expense of  \ $\phi
_{0}^{\prime }.$  In case A regular solutions terminating at a maximum of
the potential can exist only at a discrete sequence of values of \ $\phi
_{0}^{\prime }.$   \\

\textbf{\textit{Case} B}: \\
Linearizing the set $\left(\ref{Convenient setB}\right) $ we get the following equation for $\delta \phi$ :
\begin{equation*}
\delta \phi ^{\prime \prime }+S_{1\infty }\delta \phi ^{\prime }+\left( \frac{\partial ^{2}V_{\infty }}{\partial \phi ^{2}}-\frac{d_{0}-3\varkappa ^{2}\phi _{\infty }^{2}}{\left( d_{0}+1\right) \left( d_{0}+\varkappa ^{2}\phi _{\infty }^{2}\right) }S_{1\infty }^{2}\right) \delta \phi =0.
\end{equation*}
Again its nontrivial solution has the form 
\begin{equation}
 \delta \phi =Ae^{\lambda
_{+}l}+Be^{\lambda _{-}l}, 
\label{Solutions at l to infB}
\end{equation}
where $A$ and $B$ are constants of integration, and the
eigenvalues are
\begin{equation}
\lambda _{\pm }=-\sqrt{\frac{d_{0}+1}{4\phi _{\infty }}\frac{\partial V_{\infty }}{\partial \phi }}\pm \sqrt{\left( \frac{d_{0}+1}{4}+\frac{d_{0}-3\varkappa ^{2}\phi _{\infty }^{2}}{d_{0}+\varkappa ^{2}\phi _{\infty }^{2}}\right) \frac{\partial V_{\infty }}{\phi _{\infty }\partial \phi }-\frac{\partial ^{2}V_{\infty }}{\partial \phi ^{2}}}.
\label{eigenvaluesB}
\end{equation}

Eigenvalues (\ref{eigenvaluesB}) do not look as transparent as (\ref{eigenvaluesA}). However, in practice both eigenvalues (\ref{eigenvaluesB}) are complex with negative real parts. Actually, in view of (\ref{kappa_fi<1})
\begin{equation*}
\left( \frac{d_{0}+1}{4}+\frac{d_{0}-3\kappa ^{2}\phi _{\infty }^{2}}{d_{0}+\kappa ^{2}\phi _{\infty }^{2}}\right) \frac{\partial V_{\infty }}{\phi _{\infty }\partial \phi }-\frac{\partial ^{2}V_{\infty }}{\partial \phi ^{2}}<\left( \frac{d_{0}+1}{4}+1\right) \frac{\partial V_{\infty }}{\phi _{\infty }\partial \phi }-\frac{\partial ^{2}V_{\infty }}{\partial \phi ^{2}}.
\end{equation*}
For the "Mexican hat" potential (\ref{Mexican
hat}), that we use in computations (see  section \ref{Numerical analysis} below), the r.h.s. is negative at least for $d_{0}<7.$ In case B the order parameter $ \phi(l) $ approaches its limiting value $\phi_{\infty} $ with oscillations.

\subsection{\label{Boundary conditions} Boundary conditions}

The complete sets of equations determining the structure of
topological defects in the cases  A $\left(
\ref{Gamma''+...A},\ref {Beta''+...A},\ref{First integralA}\right) $ and B $\left(
\ref{Gamma''+...B},\ref {Beta''+...B},\ref{First integralB}\right) $ of vector order parameter
are of the third order with respect to three unknowns $\gamma
^{\prime },\beta ^{\prime },$ and $\phi $. The simple solutions are
determined unanimously by fixing the values of these three functions in
any regular point. The center $l=0$ is a singular point of the
cylindrical coordinate system. The condition $\phi \left( 0\right)
=0$ fulfills for both symmetries (high and broken). $\beta
^{\prime }$ is infinite at $l=0.$ For this reason we  set the boundary
conditions very close to the center, but not exactly at $l=0.$

For numerical analysis it is convenient to deal with a system of
four first order equations solved against the derivatives: $\left(
\ref{Convenient setA}\right)$ in case A, and  $\left(
\ref{Convenient setB}\right)$ in case B.

\textbf{\textit{Case} A}

The symmetry breaking potential
$V\left( \phi \right) $\ enters the equations $\left( \ref
{Convenient setA}\right) $ only via its derivative $\frac{\partial
V}{\partial \phi }.$ If we leave only the main terms in the
boundary conditions: $U=-\frac{1}{l},$ $S_{1}=\frac{1}{l}$ at
$l\rightarrow 0,$ then we loose the information about the absolute
value of the potential. The value $V_{0}=V\left( 0\right) $
appears in the next approximation. Using the expansion $\left(
\ref{Bet'=1/l+..}\right) $ of $\beta ^{\prime }$ in the vicinity
of the center and the equation $\left(
\ref{G'-B'=-e^(d0G+B)A}\right) ,$ we express $c$ via $\gamma
_{0}^{\prime \prime }$ :

\begin{equation*} c=-\left(
d_{0}-2\right) \gamma _{0}^{\prime \prime }.
\end{equation*}

To preserve the complete information about the symmetry breaking
potential one has to write the boundary conditions at
$l\rightarrow 0$ as follows
\begin{equation}
\begin{array}{c} U=\frac{1}{3}\left(
d_{0}+1\right) \gamma _{0}^{\prime \prime
}l-\frac{1}{l},\text{\quad }  S_{1}=\frac{2}{3}\left( d_{0}+1\right)
\gamma _{0}^{\prime \prime }l+\frac{1}{l},\text{\quad } \phi =\phi
_{0}^{\prime }l,\text{\quad }Z=2\phi _{0}^{\prime }.
\end{array} \label{Boundary conditionsA}
\end{equation}
The values $\gamma _{0}^{\prime \prime },\phi _{0}^{\prime },$ and
$V_{0}$ are not independent. They are connected with each other by
$\left( \ref {relation for bound condA}\right) .$

\textbf{\textit{Case} B}

In case B the potential $V\left( \phi \right) $\ enters the equations $\left( \ref
{Convenient setB}\right) $ directly. The information about its absolute value is not lost even if we use only the main approximation in the boundary conditions at $l\rightarrow 0:$
\begin{equation}
U\left( 1-\varkappa ^{2}\phi ^{2}\right) =-\frac{1}{l},\qquad S_{1}\left( 1+\frac{\varkappa ^{2}\phi ^{2}}{d_{0}}\right) =\frac{1}{l},\qquad \phi =\phi _{0}^{\prime }l,\qquad \psi =\phi _{0}^{\prime }.
\label{Boundary conditionsB}
\end{equation}

\subsection{\label{Case A analytical solution} Case A analytical solution  with $\frac{\partial V}{\partial \protect\phi }\equiv 0$}

If the potential $V=V_{0}$ does not depend on $\phi $, then it
actually plays the role of the cosmological constant $\Lambda
=\kappa ^{2}V_{0}.$ The peculiarity of the vector order parameter in case A
is that the equations $\left( \ref{Convenient setA}\right) $ loose
the information about the potential completely  if $\frac{\partial V}{\partial
\phi }\equiv 0.$ The absolute value $V_{0}$ is present only in the boundary
conditions $\left( \ref{Boundary conditionsA}\right) .$ The
equations $\left( \ref{Convenient setA}\right) $ with
$\frac{\partial V}{\partial \phi }\equiv 0$\ and boundary
conditions $\left( \ref{Boundary conditionsA}\right) $ have the
following analytic solution
\begin{eqnarray*}
U=-\frac{\sqrt{C}}{\sinh \left( \sqrt{C}l\right) }, \text{\quad } S_{1}
=\sqrt{C}\coth \left( \sqrt{C}l\right),  \text{\quad } \phi \left( l\right)
=\frac{2\phi _{0}^{\prime }}{\sqrt{C}}\tanh \allowbreak
\frac{\sqrt{C}l}{2},
\end{eqnarray*}
where
\begin{equation} C=2\left( d_{0}+1\right) \gamma
_{0}^{\prime \prime }=-\frac{2\left( d_{0}+1\right) }{d_{0}}\left(
2\kappa ^{2}\phi _{0}^{\prime 2}+\Lambda \right) .
\label{C=2d0(do+1)gamma''}
\end{equation}
The solution is regular if $C\geq 0,$ i.e. $\Lambda \leq -2\kappa
^{2}\phi _{0}^{\prime 2}.$ For $g_{00}=e^{2\gamma }$ and
$r=e^{\beta }$ we find
\begin{eqnarray*}
g_{00}\left( l\right)  =e^{2\gamma }=\left( \cosh
\frac{\sqrt{C}l}{2}\right) ^{\frac{4}{d_{0}+1}}, \text{\quad } r\left(
l\right)  =\frac{2\sinh \left( \frac{\sqrt{C}l}{2}\right)
}{\sqrt{C}}\left( \cosh \frac{\sqrt{C}l}{2}\right)
^{-\frac{d_{0}-1}{d_{0}+1}}.
\end{eqnarray*}
The inclination $\phi _{0}^{\prime }$ remains arbitrary. If $\phi
_{0}^{\prime }=0$ this solution reduces to the one found earlier
(see \cite{Bron 1} and \cite {Cline})\ for the special case $\phi
\equiv 0.$ The point is that the Einstein equations with a
negative cosmological constant have a nontrivial solution (with a
nonzero order parameter) even without a symmetry breaking
potential.

In case A a necessary condition of regular solutions with broken symmetry
is the existence of extremum points of $V\left( \phi \right) ,$
where $\frac{\partial V}{\partial \phi }=0.$ In case $V=const$ the
condition $\frac{\partial V}{\partial \phi }=0$ is fulfilled
identically, and formally the order parameter $\phi $ can tend to
any  $\phi _{\infty }$ as $\ \ l\rightarrow \infty .$ The
displayed above analytical solution shows that
the existence of a negative cosmological constant is sufficient for
the symmetry breaking of a uniform plain bulk.

The special case $C=+0$ in $\left( \ref{C=2d0(do+1)gamma''}\right)
$ when $\gamma _{0}^{\prime \prime }=0$ and  \begin{equation} \phi
_{0}^{\prime }=\pm \sqrt{-\frac{\Lambda }{2\kappa ^{2}}}
\label{Fi'0=}
\end{equation}
corresponds to the plain bulk $g_{00}\left( l\right) =1,$ and
$r\left( l\right) =l$.

\section{\label{Numerical analysis} Numerical analysis}
\subsection{\label{Regular solutions in the space of parameters} Regular solutions in the space of parameters}

The numerical integration of the sets of equations $\left( \ref{Convenient
setA}\right) $  and $\left( \ref{Convenient
setB}\right) $ is performed for the ``Mexican hat'' potential taken
in the same form as in \cite{Bron 1}: \begin{equation}
V=\frac{\lambda \eta ^{4}}{4}\left[ \varepsilon +\left(
1-\frac{\phi ^{2}}{\eta ^{2}}\right) ^{2}\right]  \label{Mexican
hat}
\end{equation}

The dimensionless parameter $\varepsilon $ moves the ``Mexican
hat'' up and down. It is equivalent to adding a cosmological
constant. The energy of spontaneous symmetry breaking is
characterized by $\eta ^{2/\left( D-2\right) },$ and
\begin{equation}
a=\frac{1}{\sqrt{\lambda }\eta } \label{scale}
\end{equation}
determines, as usual, the
length scale. In most cases $a$ is associated with the core radius
of a topological defect.  The strength of gravitational field is
characterized by the dimensionless parameter
\begin{equation}
\Gamma =\kappa ^{2}\eta ^{2}.  \label{Parameter Gamma}
\end{equation}
Without loss of generality we set in computations $a=1$ and $\eta=1$ (i.e., measure $l$ in units $a,$ and $\phi$ -- in units $\eta.$)

In both cases A and B of vector order parameter the state of broken symmetry
is controlled by four dimensionless parameters $d_{0},\varepsilon ,$ $\Gamma ,$
and $\phi _{0}^{\prime }.$ It is the main difference from the scalar multiplet case where the regular configurations with given
$d_{0},\varepsilon ,$ and $\Gamma $ existed only for some fixed values
of $\phi _{0}^{\prime }.$ Now the regular configurations with
given $d_{0},\varepsilon ,$ and $\Gamma $ exist within the whole
interval $0<\phi _{0}^{\prime }\leq \phi _{0\text{ max}}^{\prime
}.$ The upper boundary of the interval $\phi _{0\text{ max}}^{\prime }$
is a function of $d_{0},\varepsilon ,$ and $\Gamma .$ This additional
parametric freedom allows to forget about the so called ``fine
tuning'' of the physical parameters.

For visual demonstration it makes sense to fix $d_{0}=4$\ and one of
the three other parameters. Then the area of existence of regular
solutions can be presented as a map in the plane of two remaining parameters.

\begin{figure}  \centering
    \includegraphics{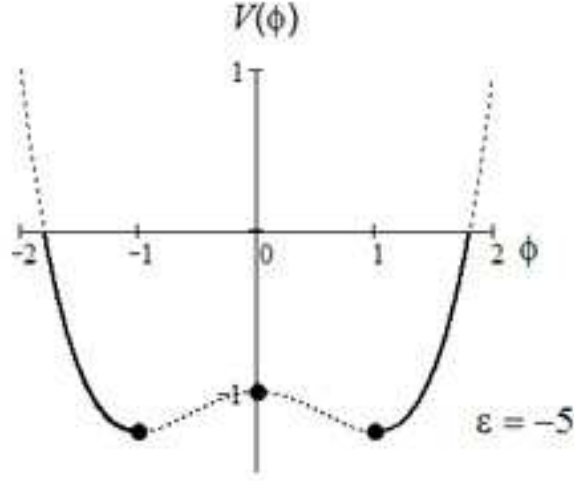}
    \caption{\label{fig:Figure 1} Mexican hat potential (dot line, $\varepsilon=-5$). In case A regular solutions terminate at the points of extremum (black points). In case B the solutions terminate on the slopes of the potential (black solid lines). }
\end{figure}

\textbf{\textit{Case} A}  \\
In case A regular solutions terminate at the points of extremum of $V(\phi).$ In dimensionless units the potential $\left( \ref{Mexican hat}\right) $ has three
extremum points -- a maximum at $\phi =0$, and two minima at
$\phi =\pm 1,$ (black points in Fig \ref{fig:Figure 1}.) At the limiting values of the order parameter
\begin{eqnarray*} V_{\infty }^{\prime } &=&0,\quad V_{\infty
}^{\prime \prime }=2,\qquad \phi _{\infty }=\pm 1 \\
V_{\infty }^{\prime } &=&0,\quad V_{\infty }^{\prime \prime
}=-1,\qquad \phi _{\infty }=0.
\end{eqnarray*}
\begin{figure}  
    \includegraphics{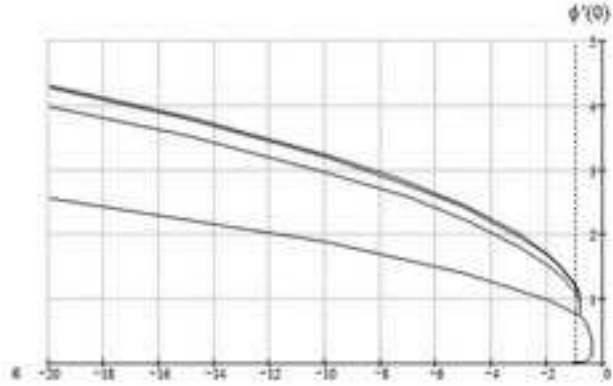}
    \caption{\label{fig:Figure 2} The area of regular configurations in the plane $\left( \varepsilon ,\phi
_{0}^{\prime }\right) $ for $d_{0}=4$ and $\Gamma =1.$ The upper curve is the boundary of existence of regular solutions. Other curves separate the regions with different signs of $\phi_\infty. $ They quickly condense to the upper curve. Below the lowest curve $\phi(l)$ does not change sign.}
\end{figure}
In case A the value of $\Gamma$ is not restricted from above.  Fig \ref{fig:Figure 2}. shows the  area of regular configurations in the
plane $\left( \varepsilon ,\phi _{0}^{\prime }\right) $ for
$d_{0}=4$ and $\Gamma =1.$ Depending on the values of $\varepsilon
$ and $\phi _{0}^{\prime }$ the order parameter $\phi \left(
l\right) $\ tends to $+1,$ $0,$ or $-1$ as $l\rightarrow
\infty .$ The sequence of curves $f_{n}\left( \varepsilon \right)
$ in Fig. \ref{fig:Figure 2} are those where $\phi \left( l\right) \rightarrow 0$
as $l\rightarrow \infty .$ They separate the areas with different
signs of $\phi _{\infty }.$ Below the first curve $f_{1}\left(
\varepsilon \right) $ from the bottom, where $0<\phi _{0}^{\prime
}<f_{1}\left( \varepsilon \right) ,$ the order parameter $\phi
\left( l\right) $ doesn't change sign. Between $f_{1}\left(
\varepsilon \right) <\phi _{0}^{\prime }<f_{2}\left( \varepsilon
\right) $ it changes the sign once. In the area $f_{2}\left(
\varepsilon \right) <\phi _{0}^{\prime }<f_{3}\left( \varepsilon
\right) $ it changes the sign twice, and so on. The curves
$f_{n}\left( \varepsilon \right) $ quickly condense to the upper
 curve $f_{\infty }\left( \varepsilon \right) $ as
$n\rightarrow \infty .$ $f_{\infty }\left( \varepsilon \right) $
is the upper boundary of existence of regular solutions (for the
particular values $d_{0}=4$ and $\Gamma =1).$

The curves in Fig. \ref{fig:Figure 2} are those where
\begin{equation}
\phi _{\infty }\left( \phi _{0}^{\prime },\varepsilon
,d_{0}=4,\Gamma =1\right) =0.
\end{equation}
\begin{figure} 
    \includegraphics{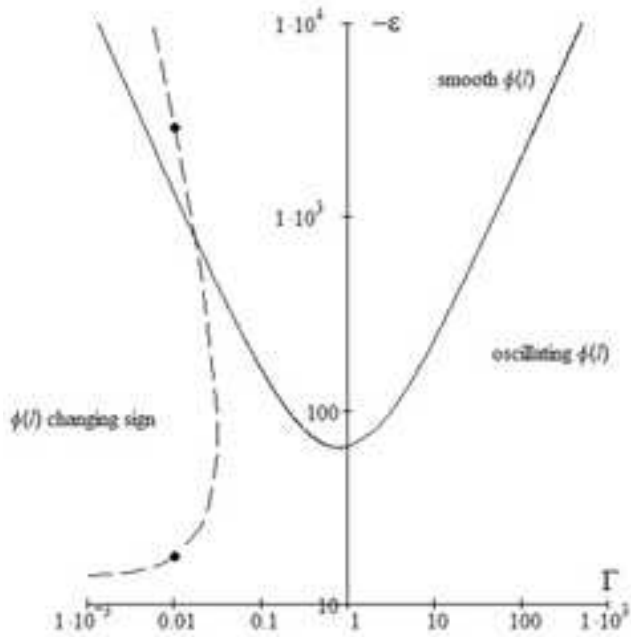}
    \caption{ \label{fig:Figure 3} Case A. Map of regular solutions in the plane
    $(\Gamma,-\varepsilon)$ for
    $\phi_{0}^{\prime}=\pm\sqrt{-\frac{\varepsilon+1}{8}}$,  $d_{0}=4.$ The solid curve (\ref{G=}) separates the regions of smooth (above) and oscillating (below) behavior of the order parameter at $l\rightarrow \infty.$ To the left of the dash curve the order parameter changes sign.}
\end{figure}
Similar curves can
be shown for fixed $\phi _{0}^{\prime }$ in the plane $\left(
\varepsilon ,\Gamma \right) .$ For instance, the dash line  in Fig. \ref{fig:Figure 3}
 is the first one of the curves $\phi _{\infty }\left( \phi
_{0}^{\prime }=\pm \sqrt{-\frac{\varepsilon +1}{8}},\varepsilon
,d_{0}=4,\Gamma \right) =0,$ where the order parameter tends to
zero at $l\rightarrow \infty $. The value $\phi _{0}^{\prime }=\pm
\sqrt{-\frac{\varepsilon +1}{8}}$ $\left( \ref{Fi'0=}\right) $
corresponds to $\gamma _{0}^{\prime \prime }=0$ in $\left( \ref
{relation for bound condA}\right) .$ It is the case $C=0$ in (\ref
{C=2d0(do+1)gamma''}), so that the symmetry breaking of the plain
bulk is caused completely by the potential $V\left( \phi \right)
,$ and not by the cosmological constant. To the right of the dash
line $\phi \left( l\right) $ does not change the sign.

\begin{figure}  \centering
    \includegraphics{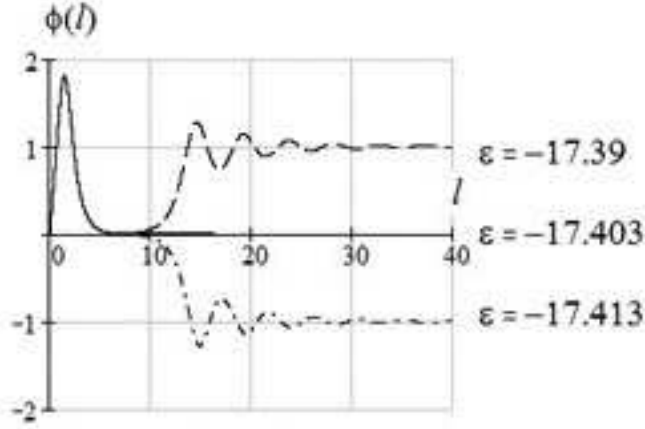}
    \caption{\label{fig:Figure 4} Oscillating solutions in the  vicinity of the lower black point on the dash curve in Fig. \ref{fig:Figure 3}.}
\end{figure}

\begin{figure}\centering
    \includegraphics{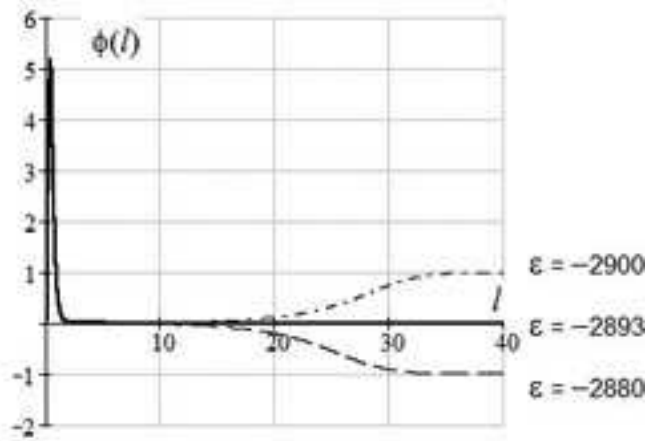}
    \caption{\label{fig:Figure 5} Smooth solutions in the  vicinity of the upper black point on the dash curve in Fig. \ref{fig:Figure 3}.}
\end{figure}

For the potential $\left( \ref{Mexican hat}\right) $ the boundary
line $\left( \ref{Boundary between mon and osc}\right) $ between
oscillating and smooth $\phi \left( l\right) $ is \begin{equation}
-\varepsilon _{b}=16\frac{\left( 1+G\right) ^{2}}{G},\quad
G=\frac{d_{0}+1}{d_{0}\Gamma }.  \label{G=}
\end{equation}
It is presented in Fig. \ref{fig:Figure 3} (solid line). Below the solid line the order
parameter $\phi \left( l\right) $ tends to its limiting value
$\phi _{\infty }$ with damping oscillations (see Fig. \ref{fig:Figure 4}), and
above this curve -- without oscillations, see Fig. \ref{fig:Figure 5}. \\
The curves
in Fig. \ref{fig:Figure 4} correspond to the close vicinity of the lower black point
on the dash curve in Fig. \ref{fig:Figure 3}, the curves in Fig. \ref{fig:Figure 5}-- to the
vicinity of the upper black point.

\begin{figure}
  \includegraphics{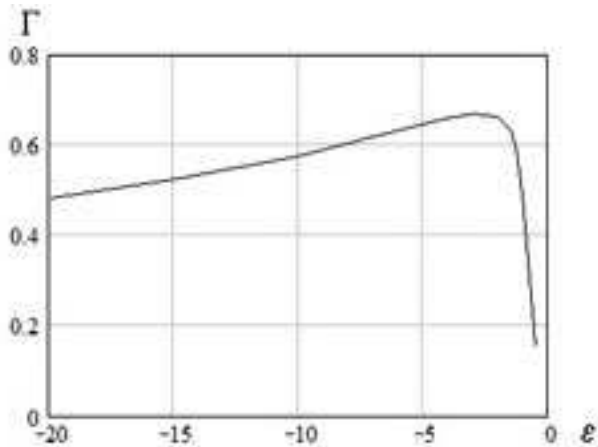}\\
  \caption{ Case B. The upper boundary of regular solutions found numerically. }\label{Figure 6}
\end{figure}

\textbf{\textit{Case} B}  \\
There are two major qualitative differences between cases A and B. In case B:

1. Parameter $\Gamma$ (\ref{Parameter Gamma}), characterizing the strength of gravitational field, is restricted from above, as it follows from (\ref{kappa_fi<1}). In the plane of parameters $(\Gamma,  \varepsilon)$ the upper boundary of existence of regular solutions, found numerically, is presented in Figure \ref{Figure 6}. 

2. Regular solutions terminate with $\phi_{\infty}$ on a slope of the potential $V(\phi),$ and not at the points of extremum.  See black solid parts on the dashed curve in Fig.\ref{fig:Figure 1},  where the conditions (\ref{V<0}) and (\ref{V'>0}) are fulfilled.

The final value $\phi_{\infty}$ obeys the equation (\ref{eq_for_fi_infB}). In case of Mexican hat potential (\ref{Mexican hat}) we find
\begin{equation}
\phi _{\infty }^{2}=1+\frac{1}{3}\left( 1+\frac{d_{0}}{\Gamma }\right) \left( \sqrt{1-\frac{3\varepsilon }{\left( 1+d_{0}/\Gamma \right) ^{2}}}-1\right).
\label{fi_inf=}
\end{equation}
If $\Gamma$ is small, (\ref{fi_inf=}) reduces to
\begin{equation}\label{fi_inf at_small Gamma}
    \phi _{\infty }^{2}=1+\frac{\Gamma \left| \varepsilon \right| }{2d_{0}},\qquad \Gamma \ll 1,
\end{equation}
and $\phi_{\infty}$ goes to a minimum of
$V(\phi)$ in the limit $\Gamma\rightarrow0$.

\begin{figure}
  \includegraphics{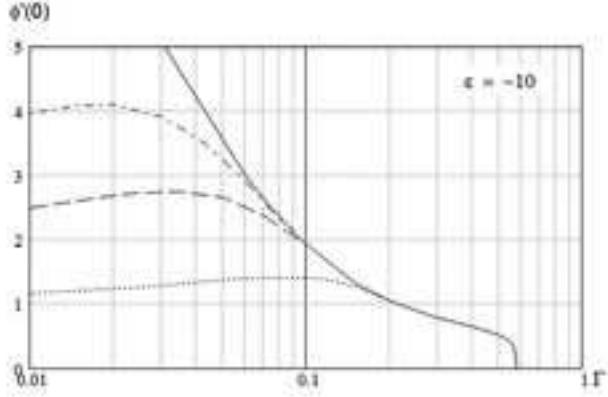}\\
  \caption{Map of regular solutions in the plane of parameters $(\phi'_{0}, \Gamma).$ Case B, $\varepsilon=-10,$ $d_{0}=4.$ The solid line is the upper boundary of existence of regular solutions. Dotted, dashed and dadot lines separate the regions where $\phi(l)$ changes its sign zero, once, twice, and three times. Separating lines between the regions where $\phi(l)$ changes sign more than three times are not shown. }\label{Figure 7}
\end{figure}

Typical map of regular solutions in the plane of parameters $(\phi'_{0}, \Gamma)$ is presented in Figure \ref{Figure 7}. Here $\varepsilon=-10,$ and $d_{0}=4.$ 
The solid line in Figure \ref{Figure 7} is the upper boundary of existence of regular solutions. The dotted line separates the solutions with the same sign of $\phi(l)$ (below) from the solutions with $\phi(l)$ changing sign ones (between dotted and dashed lines). Between dashed and dadot lines  $\phi(l)$ changes its sign twice, and so on. There also are solutions with $\phi(l)$ changing sign more than three times, but other  separating lines are not shown in the Figure \ref{Figure 7}.

\begin{figure}
  \includegraphics{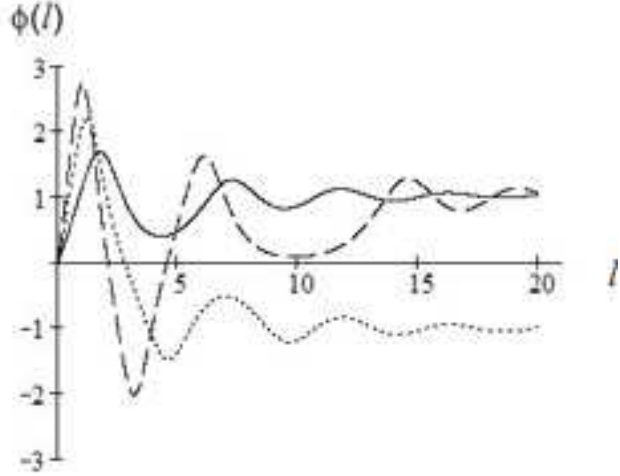}\\
  \caption{Case B. $\Gamma=0.05$ and $\varepsilon=-5$.  Order parameter $\phi(l)$ of the same sign (solid curve), changing sign once (dotted curve), and twice (dashed curve) Corresponding values of $\phi^{\prime}(0)$ are 1, 2, and 3. In accordance with (\ref{fi_inf=}) $\phi_{\infty}=\pm 1.01530$.}\label{Figure 8}
\end{figure}

The order parameter $\phi(l)$ is presented in  Figure \ref{Figure 8} for $\Gamma=0.05$, $\varepsilon=-5$, and  $\phi^{\prime}(0)$ taken from the different regions in Figure \ref{Figure 7}. Solid, dotted, and dashed curves correspond to the order parameter of the same sign, changing sign once, and twice. The values of $\phi^{\prime}(0)$ are 1, 2, and 3, respectively.

\subsection{\label{Neutral quantum particle} Neutral quantum particle in the space-time with metric (\ref{metric}) }

A neutral spinless quantum particle is described by a scalar wave function $\chi $ with the Lagrangian  \\ \begin{equation}
L_{\chi }=\frac{1}{2}g^{AB}\chi _{,B}^{\ast }\chi _{,A}-\frac{1}{2}m_{0}^{2}\chi ^{\ast }\chi .
\end{equation}
In the uniform bulk (while the symmetry is not broken) it is a free particle in the $D$-dimensional space-time with mass $m_{0}$ and spin zero. In the broken symmetry space-time with metric (\ref{metric}) it satisfies the Klein-Gordon equation
\begin{equation}
\frac{1}{\sqrt{-g}}\left( \sqrt{-g}g^{AB}\chi _{,A}\right) _{,B}+m_{0}^{2}\chi =0.
\end{equation}

The metric (\ref{metric}) depends on the only one coordinate $x^{d_{0}}=l$. So the momenta, conjugate to all other coordinates, are quantum numbers. The wave function in a quantum state is
\begin{equation}
\chi \left( x^{A}\right) =X\left( l\right) \exp \left( -ip_{\mu }x^{\mu }+in\varphi \right) ,
 \end{equation}
 where $p_{\mu }=\left( E,\mathbf{p}\right) $ is the $d_{0}$-momentum within the brane, and $n$ is the integer angular momentum conjugate to the circular extradimensional coordinate $\varphi .$ $X\left( l\right) $ satisfies the equation \cite{Bron 1}
\begin{equation}	
X^{\prime \prime }+S_{1}X^{\prime }+\left( p^{2}e^{-2\gamma }-n^{2}e^{-2\beta }-m_{0}^{2}\right) X=0. \label{X''}
\end{equation}
The eigenvalues of  $p^{2}=E^{2}-\mathbf{p}^{2}$ compose the spectrum of
squared masses, as observed within the brane. Quantum number $n$ is the
integer proper angular momentum of the particle.
From the point of view of the observer in the brane it is the internal
 momentum, identical to the spin of the particle.

The equation $\left( \ref{X''}\right) $ takes the form of the Schrodinger equation
\begin{equation}	
y_{xx}+\left[ p^{2}-\textsf{V}_{g}\left( x\right) \right] y=0
\end{equation}
after the substitution  \\

  $dl=e^{\gamma }dx,\qquad X\left( l\right) =y\left( x\right) /\sqrt{f\left( x\right) },$  \qquad
  $f\left( x\right) =\exp \left\{ -\frac{1}{2}\left[ \left( d_{0}-1\right) \gamma +\beta \right] \right\} .$ \\

The gravitational potential
\begin{equation}
\textsf{V}_{g}\left( x\right) =e^{2\gamma }\left( e^{-2\beta }n^{2}+m_{0}^{2}\right) +\frac{1}{2}\frac{1}{\sqrt{f}}\frac{d}{dx}\left( \frac{1}{f^{1/2}}\frac{df}{dx}\right)
 \label{gr.p}
 \end{equation}
 determines the trapping properties of particles to the brane. The trapping is insured by the exponentially growing warp factor $e^{2\gamma}$: in view of (\ref{Gam'_inf =A}) and (\ref{eq_for_gamma_infB}) in both cases A and B $\gamma_{\infty}^{\prime}$ are positive constants. In terms of $U$, $S_{1},$ and $\phi$  (\ref{new functions})  the dependence of the gravitational potential (\ref{gr.p}) on the distance $l$ is   \\
 \begin{equation}
 \textsf{V}\left( l\right) =e^{2\gamma }\left( e^{-2\beta }n^{2}+m_{0}^{2}\right)
  +\frac{e^{2\gamma }}{4}\frac{\left( d_{0}S_{1}-U\right) \left( U+\left( d_{0}+2\right) S_{1}\right) }{\left( d_{0}+1\right) ^{2}}+\frac{e^{2\gamma }}{2}\left( \kappa ^{2}\frac{\partial V}{\partial \phi }\phi +\frac{U\left( S_{1}-d_{0}U\right) }{d_{0}+1}\right). \label{V(l)}
\end{equation}

\subsection{\label{Oscillations} Oscillations}

\begin{figure}
  \includegraphics{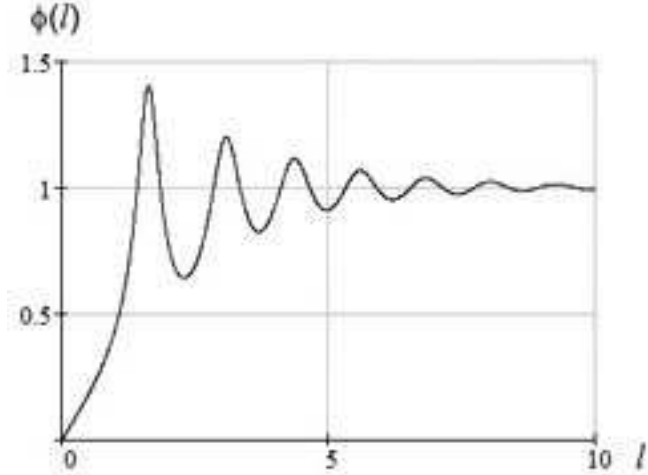}\\
  \caption{Case A. A solution with oscillating order parameter $  \phi(l).$ Here $d_{0} =4,$  $\varepsilon=-2,$ $\Gamma=10,$  $\phi^{\prime}(0)=\sqrt{-\frac{\varepsilon +1}{8}}$.}\label{Figure 9}
\end{figure}

\begin{figure}
  \includegraphics{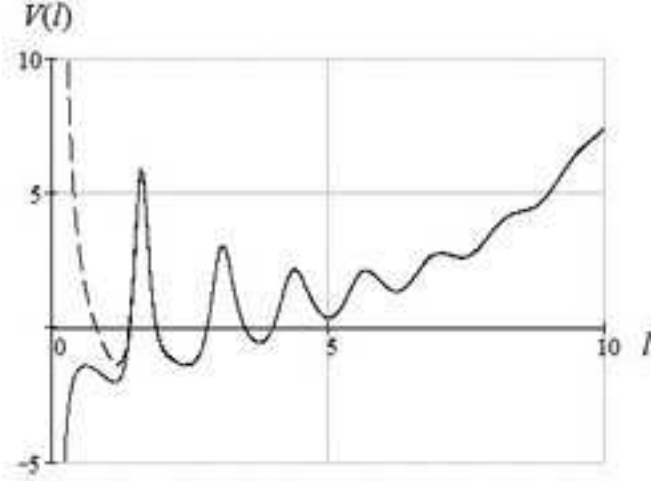}\\
  \caption{Case A. Gravitational potential $\textsf{V}_{g}\left( l\right)$ (\ref{V(l)}) for the same set of the parameters as in Fig.\ref{Figure 8}, $d_{0} =4, \varepsilon=-2, \Gamma=10, \phi_{0}^{\prime }=\sqrt{-\frac{\varepsilon +1}{8}}$. The initial mass of a test particle is set $m_{0}=0$. The solid curve corresponds to the angular momentum $n=0$, and the dashed one -- to $n=1$. }\label{Figure 10}
\end{figure}

The oscillations of the order parameter $\phi(l)$ give rise to the oscillations of the gravitational potential  (\ref{gr.p}), see Figures \ref{Figure 9},\ref{Figure 10}.

In terms of $\left( \ref{G=}\right) $\ the eigenvalues $\left(
\ref {eigenvaluesA}\right) $ are
\begin{equation} \lambda _{\pm
}=-\sqrt{-\frac{\varepsilon }{8\left( G+1\right) }}\left[ 1\pm
\sqrt{1+\frac{16}{\varepsilon G}\left( G+1\right) ^{2}}\right] .
\end{equation}
The
less is $\left| \varepsilon \right| $ the more oscillations
display themselves. In the limiting cases of small and large $\Gamma $ the
frequencies of oscillations

\begin{equation*} \left|
\textrm{Im}\lambda \right| =\left\{ \begin{array}{c} \sqrt{2},\quad
\Gamma \rightarrow 0, \\ \sqrt{2\left( 1+\frac{1}{d_{0}}\right)
\Gamma },\quad \Gamma \rightarrow \infty
\end{array}\right.
\end{equation*} \\
do not depend on $\varepsilon $ as $l\rightarrow \infty .$ At $\left| \varepsilon \right| \sim 1$ and $\Gamma \gg 1$ the gravitational potential
has many points of minimum, see Fig. \ref{Figure 10}.

\begin{figure}
  \includegraphics{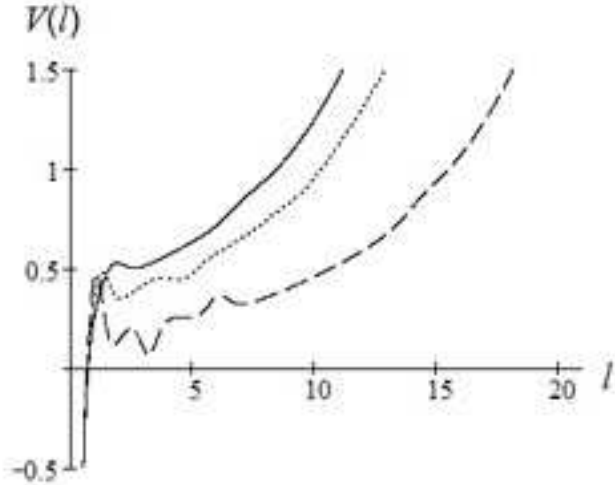}\\
  \caption{Case B. Gravitational potential $\textsf{V}_{g}\left( l\right)$ (\ref{V(l)}) for the same sets of the parameters as for the curves of order parameter in Fig.\ref{Figure 8}. $d_{0} =4$, $ \varepsilon=-5,$ $ \Gamma=0.05,$  $ \phi^{\prime }(0)=1$ (solid curve), $2$ (dotted curve), and $3$ (dashed curve). The initial mass of a test particle  $m_{0}=0$ and the angular momentum $n=0$. }\label{Figure 11}
\end{figure}
\begin{figure}
  \includegraphics{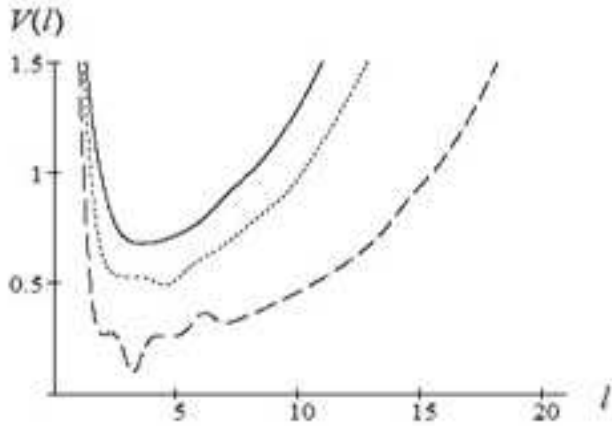}\\
  \caption{Case B. Gravitational potential $\textsf{V}_{g}\left( l\right)$ (\ref{V(l)}) for a particle with mass $m_{0}=0$ and angular momentum  $n=1$. The  sets of the parameters are the same as  in Fig.\ref{Figure 11}: $d_{0} =4$, $ \varepsilon=-5,$ $ \Gamma=0.05,$  $ \phi^{\prime }(0)=1$ (solid curve), $2$ (dotted curve), and $3$ (dashed curve). The difference between angular momenta $n=0$ and $n=1$ is essential only near the center.}\label{Figure 12}
\end{figure}

The length scale $a$ (\ref{scale}) remains an arbitrary parameter of the theory. The physical interpretation is different in the limiting cases of large and small $a.$ If $a$ is extremely large, each minimum of the potential $V(l)$ (\ref{V(l)}) forms its own brane. If the potential barrier is high, the branes are separated from one another.

In the opposite limit, when the scale length $a$ is extremely small, all points of minimum are located within one common brane, and in the spirit of Kalutza-Kline the points of minimum are beyond the resolution of modern devices.

Low energy particles can be trapped by the points of minimum of the potential (\ref{gr.p}). Identical in the bulk neutral spin-less particles,
 being trapped in the different minimum points, acquire different masses and angular momenta. If the scale length $a$ is extremely small, then they appear to
the observer in brane  as different
particles with integer spins.

\section{\label{Concluding remarks} Concluding remarks}

As a result of  rather wearing derivations of the energy-momentum tensors we get more simple equations than in the case of scalar multiplet models.

The main features of spontaneous symmetry braking with a hedge-hog type vector order parameter in comparison with the widely used previously scalar multiplet model are presented in table \ref{Table}.
\begin{table}\caption{Mutual comparison of main features of regular solutions employing the vector order parameter (cases A and B)  with the scalar multiplet model. $n_{V}$ is the number of free parameters of the symmetry breaking potential $V(\phi)$.}
  \centering
        $\begin{pmatrix}
      \textbf{Property} & \textbf{Case A} &\textbf{Case B} & \textbf{Scalar multiplet}\\
      $Order  of  Einstein  equations$ & 3 & 3 & 4 \\
      $Number of free  parameters$ & n_{V}+1 & n_{V}+1 & n_{V} \\
      $Fine  tuning$ & $\textit{no need}$ & $\textit{no need} $& \textit{in some cases} \\
      r(\infty)   & \infty & \infty & \infty, r_{m},  0\\
      $Matter trapping$  & yes & yes & yes \\
      $Presence of $ dV/d\phi $ in equations $& yes & no & no \\
     $Presence of $ V $ in  equations$  & no & yes & yes \\
     $Derivation of $ T_{IK}& wearing & wearing & easy  \\
     $Equations are$ & $\textit{most simple}$ & \textit{more simple} & \textit{simple}\\
     $Strength of gravitational field $ \Gamma& $\textit{arbitrary}$ & \textit{restricted from above} & \textit{arbitrary}\\
    \end{pmatrix}$
    \label{Table}
\end{table}
The solutions have additional parametric freedom: in both cases A and B the inclination $\phi_{0}^{\prime}$ is arbitrary within a whole interval of values (see Figures \ref{fig:Figure 2} and \ref{Figure 6}), and not only at some fixed values as in the scalar multiplet case. It means that the possibility of existence of the brane world is not connected with any restrictions of fine-tuning type. The origin of the additional parametric freedom is the order of equations, which in case of vector order parameter is less than in scalar multiplet models.

All regular configurations display trapping properties.  Oscillating behavior of the order parameter, especially in case A, gives rise to existence of several points of minimum of the attractive potential (\ref{V(l)}) at different energy levels. Particles,  trapped at different points of minimum,  acquire different masses. If we assume that the length scale  $a$ (\ref{scale}) is extremely small, it can be a reason of the observed hierarchy of masses. Angular momentum of extra-dimensional motion $n$ is also a quantum number. It appears to the observer on brane as an internal momentum of the particle which can be scarcely separated from its spin.

Most elementary particles have half-integer spins. The simple case of spontaneous symmetry breaking, considered above, can not connect the origin of half-integer spins with extra-dimensional angular momenta. Half-integer spins in General Relativity is still a problem.

\end{document}